\documentclass[aps,pr, amsfonts, twocolumn, groupedaddress]{revtex4}
\usepackage{graphicx}
\begin{document}

\title{Postulates and measurements in Everett's Quantum Mechanics}
\author{Per Arve}
\email[]{per.arve@me.com}
\address{Johan Enbergs v\"{a}g 19, 171 62 Stockholm, Sweden}

\date{\today}

\begin{abstract}
Everett's Relative State Interpretation (aka Many Worlds Interpretation) has gained increasing interest due to the progress understanding the role of decoherence. In order to fulfill its promise as an intellectually economic realistic description of the physical world, two postulates are formulated. In short they are 1) for a system with continuous coordinates $\mathbf x$, discrete variable $j$, and state $\psi_j(\mathbf x)$, the density $\rho_j(x)=|\psi_j(x)|^2$ gives the distribution of the location of the system with the respect to the variables $x$ and $j$; 2) an equation of motion for the state $i\hbar \partial_t \psi = H\psi$. The contents of the standard (Copenhagen) postulates are derived including the quantum probabilities (Born's rule).
\end{abstract}

\maketitle

\section{Introduction}
Early on, quantum mechanics was judged unable to describe the measurement process. 
The spatial spread of the wave function was in conflict with that each recorded particle was found in a well-defined place or direction. 
Born gave the rule that the probability distribution is given by the wave function absolute squared $|\psi|^2$ \cite{born1926quantenmechanik}. 
Bohr and Heisenberg took the view that quantum mechanics is correct in the microscopic world. 
However, at the act of measurement, a transition is necessary to ``classical" description for the macroscopic detector. 
This view has been called the Copenhagen Interpretation. 
It is sometimes taken to mean that the wave function collapses into a state consistent with the measured value. 
No mechanism or explanation of the transition from quantum to classical or the collapse was given.
The dichotomy between quantum and classical description has been problematic and led to statements of the kind ``quantum mechanics is impossible to understand". 

Mermin \cite{Mermin:1989aa} wrote  ``shut up and calculate" to sum up what the Copenhagen Interpretation meant to him. 
It has served us well to calculate system properties with the quantum equations and using Born's probability rule when applicable. 
Further investigations of the foundation of quantum mechanics have been given a low priority within the physics community. 
The initial acceptance of the Copenhagen Interpretation may be understood from that a partial description is better than no description of nature. 
But, a full description is more desirable than a partial description. 

Einstein refused to believe that the fundamental physics is probabilistic \cite{born1971born}. 
The quantum physics could at best be an effective theory, covering up a deeper reality. 

The wave function amplitude is sometimes called the probability amplitude \cite{dirac1967principles} as if its sole purpose is to give probabilities. 
But that terminology suggests that a classical particle is lurking behind the scenes. 
This is difficult to reconcile with the superposition principle, which is of ultimate importance. 
For the particle to be revealed there has to be a deviation from the state vector description. 
We have no evidence for a mechanism that produces any deviation from the quantum state vector description.

Quantum physics has ben enormously successful in describing the detailed physics of solids, molecules, atoms, nuclei and particle physics. 
Is it really sensible to claim that this theory only describes what happens in between preparation and measurement, but not the measurement process? 
When I sit on a chair, I do a continuous observation of its stability that is explained by the wave nature of electrons. If the observation makes quantum physics invalid, how come the chair still holds me up? 
As we do observations all the time of phenomena that are explained by quantum mechanics, it seems that quantum mechanics also describes what happens during measurements.
 
An important step towards erasing the dichotomy between state vector evolution and measurement was taken by Everett in 1957 \cite{everett1957relative}. 
In Everett's interpretation, the state vector completely describes the state of (the relevant parts of) the physical world. 
How quantum mechanics may describe the measurement process was presented. 
Everett noted that a measurement will lead to entanglement between the system being measured and the detector. 
An observer that reads of the detector can be viewed as being another detector that measures this entangled state and hence gets entangled with it. 

After the measurement, the total state is (still) a sum over all possibilities of the measured system. 
Each term describes the measured system in a state of the precise value of the measured property, the detector has registered that value so has the observer. 
These terms were called branches and they have one important property in common. 
The observer has experienced that one particular value was measured, but the observed value differs between the branches. 
The problems of previous interpretations seem to be resolved by this observation, though one important aspect was missing and another was insufficiently treated. 
Firstly, it was not shown that the branches cannot interfere which is necessary for the branching to be well defined.  Zeh \cite{zeh1970interpretation} noted this problem and realized that its solution was the process we call decoherence \cite{joos2013decoherence,zurek2003decoherence,schlosshauer2007decoherence}. This also explains why the macroscopic world we experience obeys the laws of classical physics.
Secondly, as this theory tries to describe the process of measurement in quantum mechanical terms, the probabilities as given by Born's rule should be derived from the theory. 
The latter point was attempted by Everett but his argumentation was insufficient, see section \ref{previous}.  

A weakness of Everett's theory lies in a lack of state vector interpretation. 
Everett \cite{everett1957relative} writes "The wave function is taken as the basic physical entity with {\em no a priori interpretation}. Interpretation comes after an investigation of the logical structure of the theory". This idea still prevails as illustrated by the following citation from
Tegmark  \cite{tegmark1997interpretation}  " ... postulates in English regarding interpretation would be (mathematically) derivable and thus redundant". Also in Wallace thorough exposition of Everett's quantum mechanics \cite{wallace2012emergent} the lack of interpretation of the quantities of the theory remains. 
This is criticized by Kent 	\cite{KentAgainstMWI} and Maudlin  \cite{maudlin2014critical}. The latter comments "Taking away the measurement postulates ... robs the textbook recipe of any empirical consequences".  

Equations that describe physical processes always need to be interpreted. Ballantine \cite{ballentine1973can} makes this clear in the following quote "from the formalism $f = ma$, one cannot deduce that $f$ is a force, $m$ is a mass, and $a$ is an acceleration". b-

Without an interpretation, there can be no meaning to expressions derived from the wave function.
This means there has to be: (1) a correspondence between the quantities that enter into the equations and well defined physical phenomena and observations; (2) an understanding of what the equations are able to describe, their region of applicability. 

The purpose of this article is to give an interpretation of the quantum state, suitable to describe the measurement as a quantum process including how Born's rule arises. 
For this end, quantum theory is supplied with a new set of postulates replacing the traditional postulates.

\section{Postulates \label{postulates}}

In Heisenberg's article \cite{Heisenberg:1925aa} that came to be the starting point of quantum mechanics, he aimed to replace the notion of a definite position of the electron with a quantity that could give transition probabilities using the classical dipole radiation formulas. He also aimed to reach a theory that could be generalized to more systems than the Bohr-Sommerfeld orbits could be applied to. Thus quantum mechanics is about position though the particle position concept is different from classical mechanics. 

Schr\"odinger \cite{schrodinger1926quantization} sought to find an equation for a (wave) function of space, that could give the quantized energies. At the large quantum number limit, there is a clear correspondence between wave function and the classical mechanical orbit, at least for integrable systems. Thus, the wave function replaces the where and how of the classical orbit.

\begin{description}
\item[Postulate 1 (EQM1)\label{postulateState}]  The meaning of the state:

The state is a set of complex functions of positions 
\begin{equation}
	\Psi =\{\psi_j(t,\mathbf x_1,\mathbf x_2, ...)\}
	\label{Psi}
\end{equation}
where $j$ is a discrete index, for example spin and gauge components. Its basic interpretation is given by that the density 
\begin{equation}
\rho_j(t, \mathbf x_1,\mathbf x_2,\ldots) = |\psi_j(t, \mathbf x_1,\mathbf x_2, \ldots)|^2
\label{density}
\end{equation}
answers where the system is in position, spin, etc. It is absolute square integrable normalized to one
\begin{equation}
\int\! \int \!\cdots dx_1dx_2 \cdots  \sum_j |\psi_j(t, \mathbf x_1,\mathbf x_2, \ldots)|^2 = 1.
\label{norm}
\end{equation}
This requirement signifies that the system has to be somewhere, not everywhere. 
If the value of the integral is zero, the system doesn't exist anywhere.

\item[Comment to EQM1\label{comment1}]
With the usual way of denoting the norm $ \| \cdot \| $, equation (\ref{norm}) can be written $\| \Psi \| = 1$.

If something is possible to measure, then it is possible to separate such a small part from the rest. The separated part will act as a system of its own, thus cannot have zero norm. The difference between two states $\Psi$ and $\Psi'$ for which $\| \Psi - \Psi' \| = 0$ can have no measurable consequences, as will be clear from sections \ref{singleMes} and \ref{section:repeated}. This implies that the state of the system can be viewed as a vector in the Hilbert space of functions of the type (\ref{Psi}), see the appendix. 

If the index $j$ contains gauge components, these ought to be summed over in equation (\ref{density}) to get a gauge independent density. The state function $\psi$ is not directly observable as it is gauge dependent.

\item[Postulate 2 (EQM2) \label{postulateTD}] The equation of motion:

There is a unitary time development of the state, e.g.,
\begin{equation}
i\hbar \partial_t \Psi = H \Psi,
\label{schroedinger}
\end{equation}
where $H$ is the hermitian Hamiltonian. The term unitary signifies that the value of the left hand side in (\ref{norm}) is constant for any state (\ref{Psi}) of the system.

\item[Comment to EQM2] 
When investigating how the theory describe the world we observe the Hamiltonian has to have realistic features.

The quantum world around us is understood in terms of local interactions.
The standard model of particle physics is formulated in terms of locally interacting fields.   
This implies that we only have to understand and interpret quantum mechanics with local interactions. 
In particular, measurement processes are physical processes confined to the interactions available. 
Position measurements are the single most important type of measurement, indicating that locality of interactions is a fundamental aspect of real measurements. 
In connection with measurements, it can safely be assumed that interactions are such that there is a locally conserved current. 

\end{description}

The following shortened version \cite{caves2005properties} show the essentials of the Copenhagen Interpretation: 
(C1) the state of a physical system is a normalized vector $| \Psi \rangle$ in a Hilbert space $H$ which evolves unitarily with time; 
(C2) every measurable quantity is described by a Hermitian operator (observable) $B$ acting in $H$; 
(C3) the only possible result of measuring a physical quantity is one of the eigenvalues of the corresponding observable $B$; 
(C4) the probability for obtaining eigenvalue $b$ in a measurement of $B$ is $Pr(b) = \langle \Psi |P_b |\Psi \rangle$, where $P_b$ is the projector onto the eigen-subspace of $B$ having eigenvalue $b$. 

Adding the collapse postulate "the post-measurement state in such a measurement is $P_b|\Psi \rangle/Pr(b)$" the postulates correspond to what Schlosshauer \cite{schlosshauer2007decoherence} called the standard interpretation. Some modern formulations of the postulates allow for positive operator value measurements, but that generalization offers nothing extra here. It is the same as the projection value measurement postulates (C2-4) up to a unitary transformation \cite{nielsen2010quantum}.

Both C1 and EQM1 establish that the state must be normalizable, but in the latter case, it is motivated by a physical condition.
EQM1 differs from C1 in that there is no mention of Hilbert space. 
Instead, the normalization requirement (\ref{norm}) implies that the state belongs to a Hilbert space. 

In EQM1 there is no mention of any relation between the density (\ref{density}) and probability. When the propagation of different parts is dependent on each other due to coherence, the concept of probability is not relevant. 
However,  the density $\rho_j(t, \mathbf x_1, \mathbf x_2,\ldots)$ as a distribution of the particles positions is always relevant. It is similar to Schr\"odinger's original interpretation of quantum mechanics \cite{SchrodingerIV}.

The following relations lends support to the interpretation of the density $\rho$ as the distributed position. For the sake of simplicity, the discrete index $a$, as well as the time dependence are omitted here.
\begin{itemize} 
\item[(i)] The continuity equation for a single particle,
\begin{equation}
\partial_t \rho + \nabla \cdot \mathbf j = 0,
\end{equation}
shows that the distribution of particle positions changes in a continuous manner.
\item[(ii)] In first order perturbation theory, the correction to the energy for a single particle is given by
\begin{equation}
 \Delta E = \int \! d^3x \, \rho(\mathbf x) U(\mathbf x).
\end{equation}
An outside agent interacts with the system weakly enough not to essentially change the state will find that it interacts with a distribution, not with particles in a definite position. 
\item[(iii)] We can define the average position as the first moment of the density distribution
\begin{equation}
 \langle \mathbf x \rangle = \int\! d^3x \, \mathbf x\rho(\mathbf x).
\end{equation}
According to Ehrenfest theorem, if the force $\mathbf F = \partial_x V$ is essentially constant in the region where the density is appreciable, the average position will move according to Newtons Law,
\begin{equation}
m \frac{d^2}{dt^2}\langle \mathbf x \rangle = \mathbf F
\end{equation}
If the width of the density distribution is ``small" it gives the position of a particle moving along as classical particle.
\item[(iv)] The particles are not at positions where the density is zero. 
\item[(v)] From molecular, atomic, nuclear and particle physics it is well established that the single particle density of $N$ electrons, protons or quarks
\begin{equation}
\rho(\mathbf x) = N\int d^3x_2 d^3 x_3 \cdots \rho(\mathbf x, \mathbf x_2, \mathbf x_3, \ldots )
\label{eq:single_particle_density}
\end{equation}
gives the charge density if multiplied with the charge a single particle.

In the Oppenheimer-Born approximation, the nuclei interact with the (instantaneous) charge distribution of the electrons as given by equation \ref{eq:single_particle_density}. 

In nuclear physics, the comparison between calculated charge distribution and experimental is an important method to test theories, see \cite{negele1970structure}.
\end{itemize}

As the measurement process is a physical process described by the dynamics (EQM2) there are no new postulates corresponding to C2-4. How does real physical measurement processes correspond to the C-postulates? 
In order to answer that question, this article will not address everyday observations but confine the discussion to measurements in designated experimental setups. 

For well-defined values to be recorded, it is necessary the detection system creates decoherence. 
This is a definite result of the modern analysis of Everett's quantum mechanics. If definite values can be measured in a situation where the measured state contains a variation of values, decoherence must be active. 
If not, this interpretation would be disproved.

\section{Basics of Measurements \label{singleMes}}

It is difficult to analyze, which quantities can possibly be measured based on all conceivable experimental setups. 
But, an understanding of the fundamentals of measurements can be achieved from the fundamentals of detectors.
Detectors can typically measure the position and sometimes also the kinetic energy. 
The momentum of a charged particle can be transformed into a measurement of position. 
The measurement of angular momentum in a fixed direction can be transformed into an energy measurement by the Zeeman effect or into position by a Stern-Gerlach apparatus. 
These are examples physical measurement processes that correspond to Hermitian operators. 

It is reasonable to assume that all types of measurements transform the property in question to measure a position or simply counting particles, or that the measurement procedure is related to that in the way it is calibrated. 
Thus, the following discussion of measurements will be confined to the recording of a particle entering a detector. 
This detector may be a part of an array of detectors and by that measure its position.

Particle recording detectors react when a particle is entering a certain volume or area. 
There is an infinite set of states with support inside the volume (area) and another infinite set of orthonormal states with support only outside. Together they make up a complete basis. 
The Hermitian operator that corresponds to measurements with this detector can be defined such that all the inside states are eigenstates with a common eigenvalue and the outside with another value.
This detector can only tell whether a particle came into it or not. 
A less crude position detector may be constructed by placing several such particle recorders at a multitude of positions. The Hermitian operator for this composite detector may be constructed by associating the same value for all states inside one particle recorder, but different values for different recorders. 
Additionally, another value should be attributed to the outside of all particle recorders.
This detector records if any of the individual particle recorders fired and which fired.

Obviously, the detector described so far is highly idealized. For example, it is unrealistic that a particle recording detector can register particles at any energy. 
But at a specific experiment, the energy range of the particles is limited. 
The described model is relevant as long the efficiency is close to 100\% in the real experiment. 
 
t is assumed that the measurement setup is such that, which particle recorder the particle reaches is given by its value of the property being measured. There is is a unitary operator 
\begin{equation}
U= \exp( -iH_et/\hbar)
\end{equation}
corresponding to the Hamiltonian $H_e$ that describes this part of the experimental setup.

Denote the Hermitian operator that corresponds to the position detector with $Y$. 
The operator $A\,$ being measured by $Y$ and the unitary evolution $U$ is given by 
\begin{equation}
A = U^{\dagger} Y U .
\label{Adef}
\end{equation}
The eigenstates $|a\rangle$ of $A$ are related to eigenstates of $Y$ by
\begin{equation}
|y\rangle = U |a\rangle.
\label{astates}
\end{equation}
As described above, each eigenvalue of $Y$ is typically degenerate. 
According to (\ref{Adef}), the same applies to $A$ but as noted above in an actual experiment only a small number of states are involved. For simplicity, 
it is assumed that only one state per particle recorder is relevant.

\begin{figure}[htbp]
\begin{center}
\includegraphics{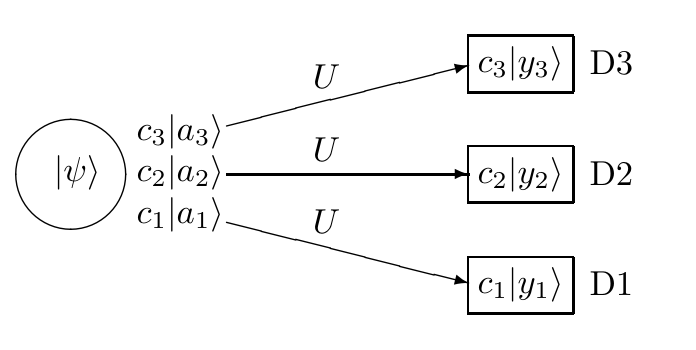}

%
%
%
%
%
%
%
\caption{The position detector consisting is the particle recorders D1-D3 receive the different components of the wave function $|\psi \rangle$ due to the unitary transformation $U$. The state $| a_n \rangle$ transforms to $| y_n \rangle$ by $U$.}  
\label{experiment}
\end{center}
\end{figure}

Sometimes, the quantity given by the operator $A$ is a vehicle to make it possible to measure another property. 
For example, in a Stern-Gerlach experiment, the physics of half-integer spins is investigated but it is the position of silver atoms that are measured.
Denote by $B\,$ be the quantity that the experiment is intended to measure. 
The eigenstates of $B$ are used as a basis for the state to be measured,
\begin{equation}
 |\psi\rangle = \sum_b c_b | b \rangle.
 \label{psi}
\end{equation}
The operators $A$ and $B$ may be the same so that $|a_b\rangle = |b\rangle$ or 
\begin{equation}
|a_b\rangle = |b\rangle \otimes |\phi_b \rangle 
\end{equation}
where the factor $|\phi_b \rangle$ is there to complete the state $| b \rangle$ to the physical particle in state $| a_b \rangle$. The state that enters the position detector system is
\begin{equation}
\sum_b c_b U|a_b\rangle = \sum_b c_b |y_b\rangle.
\label{detectordist}
\end{equation}
This expresses that the different eigenstates $|b\rangle$ enters separate particle recorders and is there represented by $|y_b\rangle$, see figure \ref{experiment}. 
Due to that the functions $y_b(j,\mathbf x)$ have disjoint spatial support, the density of the state (\ref{detectordist}) is
\begin{equation}
\rho_j(\mathbf x) =  \sum_b  |c_b|^2 |y_b(j,\mathbf x)|^2 .
\end{equation}
 
It describes where the system is according to EQM1. Summation over the spin and integration over the volume of one of the particle recorders will give the value $|c_b|^2$, where $b$ is the eigenvalue of $B$ associated with that recorder. 
The interpretation of this result is that 
\begin{equation}
\rho_b = |c_b|^2
\label{rhob}
\end{equation}
as a function of the discrete variable $b$ tells where the system is with respect to the eigenvalue of $B$. Note that this result is an important step towards replacing the old axioms C2 and C3 with EQM1 and EQM2. 

So far, the interaction between the particle and the detector has been ignored. 
The decoherence necessary for a measurement to happen relies on this interaction. 
That the interactions are local is an important assumption for the theory of decoherence and is also important to guarantee that position measurements are possible.

To simplify the discussion it will be assumed that the state $|\psi\rangle$ (\ref{psi}), rather than $|y_b \rangle$, directly interacts with the detector as the distinction between $ | b\rangle $ and $ | a_b \rangle $ is no longer needed.
Then, the interaction with the detector $M$ leads to the transition
\begin{equation}
\big(\sum_b c_b | b \rangle \big) |M_\phi \rangle  \rightarrow \sum_b c_b | b \rangle'  |M_b \rangle.
\label{eq:measurement}
\end{equation}
The  detector changes its state from its nothing registered state $ | M_\phi \rangle $ to a state $ | M_b \rangle $ consistent with having registered the state $ | b \rangle $. 
The state of the system before the measurement $ | b \rangle $  and after $ | b \rangle' $, may be the same. 
After the experimentalist has observed the detector its state still belongs to the set of states consistent with that the state $ | b \rangle $ has been registered. The observation process is described by
\begin{equation}
\big(\sum_b c_b | b \rangle'  |M_b \rangle \big) |O_\phi \rangle  \rightarrow \sum_b c_b | b \rangle'  |M_b \rangle |O_b \rangle,
\label{eq:observation}
\end{equation}
where the state of the detector $ |M_b \rangle $ is not altered. 

Due to the complicated nature of the detector, the observer and their interactions with the environment, the possibility of interference between the different terms in the final state of (\ref{eq:measurement}) or (\ref{eq:observation}) disappears rapidly. 
The different terms will belong to different ``worlds" as they will evolve independently of each other. 
This is the process of decoherence and is an essential point of the many world understanding of quantum physics. 
It guarantees that in any of the ``worlds", the observer has seen one particular value and has no direct knowledge of the readings made in the other ``worlds". 

Zeh \cite{zeh1997achieved} has pointed out that the decoherence theory depends on Born's rule, which means that the density $\rho_{\cdot}(\cdot)$ is used to interpret some results. 
As the EQM1 gives an interpretation of $\rho_{\cdot}(\cdot)$ it replaces Born's rule as a means to interpret such results. 

So far, it is clear that the contents of C2 and C3 are fully understood as a result of the unitary quantum evolution. 
The quantity being measured is associated with a Hermitian operator and the result of the measurement is one of its eigenvalues. 

Looking at the many ``worlds" from the outside the question: What reading did the observer get? is equivalent to What is the distribution of observer readings? - 
The answer is given by the distribution $\rho_b$ (\ref{rhob}). 
This value can also be arrived at calculating the total density (the norm) of  the $b$-term in the final state of (\ref{eq:measurement}) or (\ref{eq:observation}).  Note that once decoherence has taken place the created ``worlds" evolve independently, which keeps their norms conserved. 
 
\section{Repeated measurements\label{section:repeated}}

Suppose the detector is able to record several subsequent measurements of identically prepared systems (\ref{psi}). Further, assume that the way the detector interacts with the next system is not affected by previous measurements. The second measurement is described by the transition 
\begin{eqnarray}
\big(\sum_{b_2} c_{b_2} | b_2 \rangle \big) \sum_{b_1} c_{b_1} | b_1 \rangle'  |M_{b_1} \rangle \rightarrow \nonumber \\ 
\sum_{b_1b_2}  c_{b_2} c_{b_1} | b_2 \rangle' | b_1 \rangle' |M_{b_1b_2} \rangle.
\end{eqnarray}
When the interaction with the observer is included the final state becomes
\begin{equation}
\sum_{b_1b_2}  c_{b_2} c_{b_1} | b_2 \rangle' | b_1 \rangle' |M_{b_1b_2} \rangle |O_{b_1b_2} \rangle.
\end{equation}
Each sequence of readings belong to different ``worlds". The distribution of observer reading sequences is now\begin{equation}
\rho_{b_1b_2} = |c_{b_1}|^2|c_{b_2}|^2.
\end{equation}
After $N$ measurements, the sequences of observer readings are distributed according to
\begin{equation}
\rho_{b_1b_2...b_N} = |c_{b_1}|^2|c_{b_2}|^2\cdots |c_{b_N}|^2.
\label{eq:rho_rep}
\end{equation}
When $N$ is large, the relative frequencies of the values of $b$ became interesting. 
To focus on the value $b = u$, denote the summed density of all the other values of $b$ by 
\begin{equation}
\rho_{\neg u} = \sum_{b\neq u} |c_b|^2
\end{equation}
and $\rho_u = |c_u|^2$. The sum of the densities (\ref{eq:rho_rep}) over all sequences where $b=u$ appears precisely $m$ times out of $N$ measurements
\begin{equation}
\rho(m: N\, |\,u)  = \frac{N!}{(N-m)!m!}(\rho_u)^m (\rho_{\neg u})^{N-m}.
\label{eq:a_ m-times}
\end{equation}
This gives the total summed density of the worlds in which the observer has found the value $u$ $m$ times. 
Hence, the question 'how many times have the observer measured the value $u$' is answered by $\rho(m: N\, |\,u) $ as a distribution over $m$-values. 

For large number of measured systems $N$, the distribution (\ref{eq:a_ m-times}) may be approximated by a gaussian, see Feller \cite{feller1968introduction},
\begin{equation}
\rho(m: N\, |\,u) \approx \frac{1}{(2\pi N\rho_u\rho_{\neg u})^{1/2}} \exp\big(-\frac{(m-N\rho_u)^2}{2N\rho_u\rho_{\neg u}}\big).
\label{mGaussian}
\end{equation}
The distribution (\ref{mGaussian}) may be represented as function of the relative frequency $z=m/N$ taken as a continuous variable. 
The properly normalized distribution with respect to $z$ is
\begin{equation}
\rho(z | u) = \big( \frac{ N }{ 2\pi \rho_u\rho_{\neg u} } \big)^{1/2} \exp\big(-\frac{ N(z-\rho_u)^2 }{2 \rho_u\rho_{\neg u} }\big).
\label{eq:relative}
\end{equation}
As $N \rightarrow \infty$ this density approaches the delta function $\delta(z-\rho_u)$. 
This says that at infinitely large $N\,$ there is only one value of the frequency $z=\rho_u$. 
This might look like as a proof of Born's probability rule, but $\rho(z | u)$ is an approximate result. 

To get from the exact expression for $\rho(m: N\, |\,u)$ (\ref{eq:a_ m-times}) to the continuous frequency distribution, the interval $[0,1]$ is divided into a set of intervals $\{I_k\}$,
\begin{equation}
I_k = [0,1] \,\cap [z_k-\Delta z/2, \, z_k+\Delta z/2[ , \, z_k = \rho_u + k\Delta z.
\end{equation}
The index $k$ belongs to the minimal set of integers such that $\{I_k\}$ covers $[0,1]$. 
Define $\tilde{\rho}(k)$ as the sum of densities $\rho(m: N\, |\,u)$ with $m/N$ in the interval $I_k$. Let 
\begin{equation}
\rho_{\Delta z}(z|u) = \tilde{\rho}(k)/\Delta z\; \mbox{if}\; z \in I_k
\end{equation}
This is a histogram type piece-wise constant function. 
If $\Delta z = \Delta z_1/N^{-1/2}$ and $\Delta z_1$ is small and $N$ is large, then $\rho_{\Delta z}(z|u)$ can be arbitrarily close to $\rho(z | u)$. 

In order to properly justify the use of the frequency distribution (\ref{eq:relative}) an operator should be found that is closely related to this distribution.
The first guess may be the frequency operator 
\begin{equation}
F_N  = \frac{1}{N} \sum_{i=1}^N f_i 
\label{eq:frequency_operator}
\end{equation}
where $f_i$ operates on the $i$-th system being measured with $f|u\rangle=|u\rangle$ and $f|b\rangle = 0$ if $b\neq u$. 
As the operator $F_N$ is diagonal in the $| b \rangle$ basis, the density distribution of its eigenvalues for the state
\begin{eqnarray}
| \Psi \rangle = \sum_{b_1b_2\cdots b_N}  c_{b_1}c_{b_2}\cdots c_{b_N} |b_1\rangle_1|b_2\rangle_2 \cdots |b_N\rangle_N \otimes \nonumber\\
| M_{b_1b_2...b_N} \rangle |O_{b_1b_2...b_N} \rangle
\end{eqnarray}
is the same as the density distribution of 
\begin{equation}
|\psi \rangle^N = |\psi\rangle_1|\psi\rangle_2 \cdots |\psi\rangle_N 
\end{equation}
where $|\psi \rangle_i$ is the $i$-th measured system state given by (\ref{psi}). 
The eigenvalues of $F_N$ are $z = m/N, \, m= 1,...,N$. 
The density related to $F_N$ acting on this state is given by (\ref{eq:a_ m-times}) with $m$ replaced by $zN$. 
As pointed out by Squires \cite{SquiresOnProof}, this is a discrete distribution and each of its values approaches zero as $N \rightarrow \infty$. 

The operator $F_{N\Delta z}$ defined by its action on products of eigenstates to the operator $B$. If the frequency of the eigenvalue $u$ is in the interval $I_k$ with midpoint $z_k$, then
\begin{equation}
F_{N\Delta z} |b_1\rangle_1|b_2\rangle_2 ... |b_N\rangle_N = z_{k} |b_1\rangle_1|b_2\rangle_2 ... |b_N\rangle_N.
\end{equation}
The density of this operator is $\tilde{\rho}(k)$. As the eigenvalues $z_k$ of $F_{N\Delta z}$ is a discrete set its density distribution $\rho_{z_k} = \tilde{\rho}(k)$ is represented be a bar graph rather than the histogram that represents $\rho_{\Delta z}(z|u)$. 

To see the behavior of these densities as $N$ approaches infinity, the Chebyshev's  inequality \cite{feller1968introduction} can be applied to the distribution $\rho(m: N\, |\,u)$ (\ref{eq:a_ m-times}). 
The result can be written as  
\begin{equation}
\sum_{|m/N-\rho_u| > \Delta z/2} \rho(m: N\, |\,u) \leq \frac{4\rho_u\rho_{\neg u}}{\Delta z^2 N}.
\end{equation}
From this follows that $\sum_{k \neq 0}\tilde{\rho}(k) \rightarrow 0$ as $N \rightarrow \infty$ and that $\tilde{\rho}(0)$ approaches one for any value of $\Delta z$. 
This shows that the delta function limit of $\rho(z | u)$ is confirmed by the exact calculation.

\section{Born's rule \label{bornsrule}}

\begin{figure}[!htb]
\begin{center}
\includegraphics[scale=0.45]{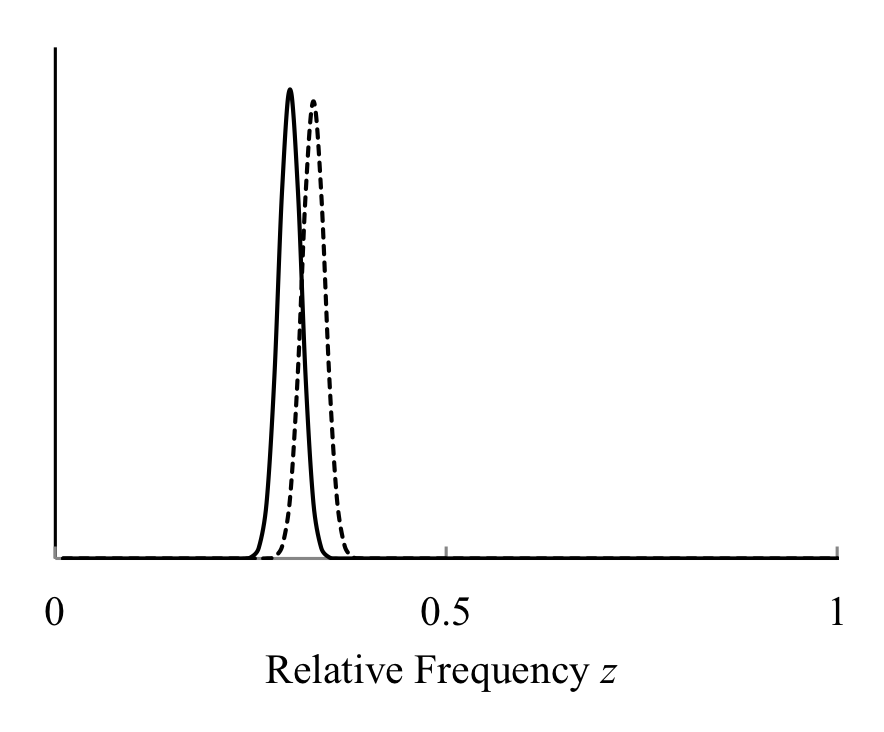}
\caption{\label{fig:binomials}The solid line shows the density  $\rho(z | u)$ (\ref{eq:relative}) for $\rho_u = 0,\!3$ and $N=1000$. 
The dotted line shows where an observer in a typical branch may estimate the probability $P(z | u)$ to be from the observed sequence alone.}
\end{center}
\end{figure}

Consider an observer who is involved in a deliberate measurement process of a phenomenon where the outcome is uncertain due to the observed quantum state containing more than one possible value. After a long sequence of measurements, the observer is distributed over very many branches. In each branch, a random sequence is observed which call for a statistical analysis by the observer. The observer will assume that there is a probability $P_u$ for measuring $u$ in a single measurement. The probability of the measured relative frequency $z$ after $N$ repeated measurements, for this value of $P_u$, is
\begin{equation}
P(z | u) =  \big( \frac{ N }{ 2\pi P_u(1-P_u)} \big)^{1/2} \exp\big(-\frac{ N(z-P_u)^2 }{ 2P_u(1-P_u) }\big).
\label{eq:Prelative}
\end{equation}
As this a very narrow distribution for large $N$, this shows that $P_u$ is probably close to $z$.
The relative frequency $z$ is distributed over all branches according to (\ref{eq:relative}), see figure \ref{fig:binomials}. Hence, the distribution of $P_u$ over the branches may be seen as the folding of the two distributions.

As the number of repeated measurements $N$ grows the width of the distribution of $P_u$ tends to zero as does $\rho(z|u)$. 
After a large number, $N$ of repeated measurements the observer sees a relative frequency close to $\rho_u$ and the large value of $N$ implies that the value of $P_u$ is probably close to the observed frequency. Hence, the observer believes that the probability $P_u$ is very close to $\rho_u$.

There is one additional effect that can make the difference between $\rho_u$ and $P_u$ go away. That happens if the observer knows what quantum state the system is in. When the observer sees a relative frequency close to $\rho_u$ the observer will likely assume that $P_u$ is precisely $\rho_u$, as stated by Born's rule. This might well happen even if Born's rule is not known to the observer, as inventing Born's rule is known to be possible. 

To summarize the present analyses of probabilities in quantum mechanics: the observer distribution in relative frequency (\ref{eq:relative}) is narrowing in precisely the same fashion as for a classic probability (\ref{eq:Prelative}). 
If it is assumed that the observer's branch is reasonably typical, the observer experiences Borns's rule when the number of repeated experiments are large, unless an exceptional sequence appears. 
Such sequences are also a consequence of Born's rule as any mechanism giving Born's rule must also give rise to possible but improbable sequences. 
In any statistical analysis it is assumed that what will be or is observed is reasonably typical, here this applies to both $P(z|u)$ and $\rho(z|u)$. 
This proves that in a typical world physicists will believe in Born's rule.

Is Born's rule proven by this? Is it motivated by the reasoning above to claim that the probability for a measuring the value $u$ is $\rho_u$? Is the word probability appropriate to use when all alternatives happen? 
Perhaps, the most convincing argument for using the concept of probability is that it (typically) will work well for the physicist who applies an appropriate probability model to existing data or for planning new experiments.

In classical probability considerations, it is sometimes argued that the use of probability is related to a lack of knowledge. 
In many-worlds quantum theory we know what is going to happen, all alternatives with non-zero amplitudes are going to be realized. 
To see how a lack of knowledge can enter here, consider the sleeping pill scenario. 
There, the observer has been asleep during the measurement \cite{vaidman1998schizophrenic}. 
When the observer wakes up she asks herself, what will I see if I look at the detector. 
The point is that in a definite branch there is a definite outcome. 
The observer is ignorant about the outcome so this situation fulfills the ignorance requisite of classical probability.

As argued by the philosophers of physics Saunders \cite{saunders2010chance}, Wallace \cite{wallace2010prove,wallace2012emergent}, Greaves and Myhrwold \cite{greaves2010everett}, the concept of classical probability concept is problematic but to some extent it is less problematic in the many-worlds theory.
The frequentist solution is to define probability from a measure on sets as described in \cite{feller2008introduction}.
Infinite sequences of events show the same frequency except for a set of sequences of measure zero.
Probability is here identified with frequency.
One critique against this view is that the calculation of frequency requires knowledge of the probability so frequency cannot be used to define it. 
The alternative is the bayesian theory \cite{jaynes2003probability} where probabilities on a fundamental level are taken to be subjective beliefs. 
Probability is here related to decisions an agent is willing to make \cite{savage1972foundations}. 

The derivation of quantum probabilities given here has much in common with the bayesian view. It is derived that physicists in a typical branch believe in Born's rule. This is in accordance with probabilities as subjective beliefs.  As in Jaynes treatise \cite{jaynes2003probability} there is a preceding quantity used to derive probabilities, which here corresponds to the location of the system. 

The present derivation may also be acceptable from a frequentists point view due to the focus on relative frequency. In the next section, it will be also shown that the state describing infinitely many repeated measurements has a well-defined frequency. Hence, in quantum physics, the frequency can be calculated without using the probability concept, unlike the classical situation. 

\section{Previous attempts to derive Born's rule \label{previous}}

To get a solid argumentation for Born's probability rule, one cannot assume the properties of the rule beforehand. 
One cannot even assume that probability applies to quantum mechanics. 
It could be that most sequences that an observer sees don't follow the patterns implied by probability laws. 
If probability applies, it may depend on the environment, previous history and more. 
Every statement should either serve as a postulate or be derived from reasonable postulates. 

In this section, several attempts to prove Born's rule in ways that could suit EQM are discussed with respect to what is proved and what is explicitly or implicitly assumed. The discussion of which unproved statements could be made into postulates is deferred to the next section (\ref{sec:altposts}).

Everett \cite{everett1957relative} shows that the observer in a repeated measurement will see a random series of results in each branch. 
He then concludes that ``we must put some sort of measure (weighting) in the elements of a final superposition". 
Everett assumes that the measure only depends on the (absolute) value of the amplitudes $c_b$ (\ref{psi}), which implies that the measure is independent of other properties of the state.
Further, he assumes the measure of a state should be the sum of measures of its orthonormal basis states. 
By this, he arrives at the measure that corresponds to Born's probability rule. 
He argues that the measure is the desired probability measure from that it is conserved over time and has the right mathematical properties for branching worlds.
Everett's `proof' is only indicative and not conclusive. 
The assumptions are reasonable but that is not enough to constitute a proof. 

Finkelstein \cite{finkelstein1963section} and Graham \cite{graham1973measurement} have tried to derive Born's rule using arguments that in effect are based on the narrowing of the relative frequency distribution (\ref{eq:relative}).
They show that in the limit $N \rightarrow \infty$  the `variance´ of the frequency operator $F_N$  (\ref{eq:frequency_operator}), $(\Delta_N F_N)^2 = ^N\langle \psi | (F_N-\rho_u)^2  |\psi \rangle^N$, tends to zero. 
As the Born rule postulate C4 have been abandoned, $\rho(z | a)$ is left without interpretation which implies $(\Delta_N F_N)^2$ has no interpretation either. To let the Born rule give it meaning, would result in a circular proof  \cite{caves2005properties,Farhi1989368,weinberg2015lectures}. Note that, the procedure of taking the limit $N \rightarrow \infty$ only deals with quantities at finite $N$ values. Without a physical interpretation of $(\Delta_N F_N)^2$ at finite $N$ the limiting procedure will have no physical significance.

To prove Born's rule, Hartle \cite{hartle1968quantum}, DeWitt \cite{dewitt1973many}, Farhi, Goldstone, Gutmann \cite{Farhi1989368,gutmann1995using} start from C1-3. The postulate C4 that contains Born’s rule have been replaced by: (C4') When the system state is an eigenstate to the operator B corresponding to the property to be measured, the measurement result will with certainty be the eigenvalue to the operator.

They take a frequentist approach and prove that
\begin{equation}
| \psi \rangle^{\infty} = \prod_{i=1}^{\infty} | \psi \rangle_i
\label{inftyrep}
\end{equation}
is an eigenstate to the corresponding frequency operator, 
\begin{equation}
F_{\infty} = \lim_{N\rightarrow \infty}F_N.
\end{equation}
DeWitt admitted that his proof was incomplete and referred to Hartle  \cite{hartle1968quantum}. 
The state (\ref{inftyrep}) belongs to the Hilbert space $H^{\otimes \infty}$, the tensor product of infinitely many single system Hilbert spaces.  The proofs are complicated as $H^{\otimes \infty}$ is non-separable, see the appendix.

Hartle considers the limited frequency operator $F_N$ that only acts on the first $N$ states of $| \psi \rangle^{\infty}$ and proves that 
\begin{equation}
\| F_{\infty}| \psi \rangle^{\infty}  - \rho_u | \psi \rangle^{\infty} \| = \lim_{N\rightarrow \infty} \| (F_N - \rho_u)| \psi \rangle^{\infty} \| = 0.
\end{equation}
Thereby, Hartle proved that $F_{\infty} $ is a well defined operator on the relevant kind of states and that $| \psi \rangle^{\infty}$ is an eigenstate with eigenvalue $\rho_u$.
Farhi, Goldstone, and Gutmann  \cite{Farhi1989368} used a basis consisting of states that represent a particular sequence of measurements.  
To get non-zero matrix elements with $| \psi \rangle^{\infty}$ the basis states were given an infinite norm. 
Caves and Schack \cite{caves2005properties} criticized that these basis states dom't belong to the Hilbert space 
causing a lack of rigor in the proof.

Gutmann \cite{gutmann1995using} subsequently produced a proof that circumvents the renormalization procedure. 
He used the mathematical similarity between the contribution to the norm of $| \psi \rangle^{\infty}$ from particular sequences of `measurement results' and the probability of that outcome in the corresponding classical calculation. 
The total classical probability of sequences with the relative frequency equal to the probability is one. 
The relations between probability, density and norm give that the total norm equals one of the part of $| \psi \rangle^{\infty}$ that consists of sequences with relative frequency $\rho_u$. 
The simplicity and elegance of this proof hinge on the reuse of probability mathematics.

Caves and Schack \cite{caves2005properties}  criticized all three treatments  \cite{hartle1968quantum,Farhi1989368,gutmann1995using} for being arbitrary. It is claimed, if another than the Hilbert space norm $\|\cdot \|$ was used, the frequency of value $u$ will not be $\rho_u$. This argument doesn't constitute a valid criticism. As it was postulated C1 that the state belongs to Hilbert space it would have been very inconsistent to use another norm. It is not even correct that another norm necessarily would have changed the result \cite{van2006many}.

Several articles \cite{Farhi1989368,cassinello1996probabilistic,caves2005properties,wada2007derivation} have not understood the fundamental difference between Hartle's derivation and that of Graham and Finkelstein. The sequence $F_N | \psi \rangle^{\infty}$ is convergent in the space $H^{\otimes \infty}$, while $F_N | \psi \rangle^N$ is a not a Cauchy sequence in Fock space.

Additionally, Cassinello and S\'{a}nches-G\'{o}mez \cite{cassinello1996probabilistic} didn't recognize that convergence and eigenvalue equations are to be evaluated using the Hilbert space norm, see the appendix. 
Squires \cite{SquiresOnProof} argued from that the quantities $\rho(m: N\, |\,u)$ (\ref{mGaussian}) vanish as $N \rightarrow \infty$ and erroneously conclude that states of definite frequency are orthogonal to $|\psi \rangle^{\infty}$. This is rather a consequence of that the states of definite frequency makes up an uncountable basis in $H^{\otimes \infty}$. 

Hartle \cite{hartle1968quantum} and Gutmann \cite{gutmann1995using} prove that the state $| \psi \rangle$ is an eigenstate to the frequency operator with the expected eigenvalue. 
Weinberg \cite{weinberg2015lectures} calls the extension of quantum calculations to the space $H^{\otimes \infty}$ "a stretch" but there are no compelling arguments agaisnt the two proofs, which are both valid and elegant. 
This shows that in the $N=\infty$ limit where $\rho(z | u)$ is a delta function, it truly represents the frequency distribution of the limiting state $|\psi \rangle^{\infty}$. 
It reassures the consistency of the reasoning in sections \ref{section:repeated} and \ref{bornsrule}.

Hartle ends his proof of Born's rule with stating that $| \psi \rangle^{\infty}$ describes an infinitely large ensemble. 
The state of the ensemble is fully coherent before the imagined measurement of the relative frequency. 
But, the individual systems in $| \psi \rangle^{\infty}$ don't interact with each other.
Thus, the calculation of the relative frequency also applies to the measurement of the systems one by one. 
But, to infer a value to the probability of measuring a particular value requires that it has been established that the values appear randomly. 
This was never done by Hartle \cite{hartle1968quantum}, Farhi et al.  \cite{Farhi1989368}, and Gutmann  \cite{gutmann1995using} so they never really proved Born's rule. 
The difficulty in deducing probability from infinite sequences is an important part of the bayesian criticism against frequentism \cite{caves2005properties}. 

Due to the conceptual difficulties with the frequentist view of probability and the related criticism against proving Born's rule with frequency arguments, other routes have been pursued. 
Following the vision that EQM needs no interpretation Deutsch \cite{deutsch1999quantum}, Zurek \cite{zurek2003environment,BornRuleEnvariance,QDarwinPhysToday}, and Sebens and Carroll \cite{sebens2014self,carroll2014many} used a variety of symmetry arguments to show that states with equal magnitude of the amplitude have equal probability. The main problem with these derivations is that without any interpretation it is impossible to argue about the properties of states, including how they are related to probabilities. To get to Born's rule they had to make a variety of assumptions. Thus their analysis don't constitute proofs but can be viewed as indicators how the probabilities may appear due to branching of worlds.

For unequal magnitudes, they assumed that 
\begin{equation}
| \psi \rangle = c_1 | 1 \rangle + c_2 | 2 \rangle, \qquad | c_1 |^2 / | c_2|^2 = p/q,
\end{equation}
where $p$ and $q$ are integers. The state $|1\rangle$ is divided into $p$ substates, and $|2\rangle$ into $q$ substates. such that $|\psi\rangle$ equals $n=p+q$ different states with equal amplitudes, rendering the probability of $| 1 \rangle$ to $p/n$ and $ | 2 \rangle$ to $q/n$. 
This approach relies on the possibility to split a state into several. The discussions have to be abstract in lieu of an interpretation of the states. This makes it impossible to consider any justification for the extra degrees of freedom assumed to exist to split the states $|1 \rangle$ and $| 2 \rangle$ into several states.
One trivial shortcoming is that $ |c_1 |^2 / |c_2|^2$ has to be rational number. But, if the proofs would work in all other respects, this limitation could be circumvented in a way similar to how Wallace \cite{wallace2012emergent} argued in one of his derivations of probabilities.

Deutsch \cite{deutsch1999quantum} used a decision theoretic starting point to arrive at equal probabilities for equal absolute amplitudes $|\alpha|=|\beta|$. 
Either state $| 1 \rangle$ or state $| 2 \rangle$ is measured. 
The ``player" get rewarded by the amount $x_1$ or $x_2$ respectively. 
The ``banker" will get $-x_1$ or $-x_2$. Using a symmetry between banker and player and by adding or subtracting the same amount Deutsch attempts to prove a symmetry between the two states. 
This procedure was very arbitrary and cannot be used to prove Born's rule.

Zurek \cite{zurek2003environment,BornRuleEnvariance,QDarwinPhysToday} argued that there is an ``environment assisted of invariance" called envariance, which is used to establish that equal magnitudes imply equal probability. 
The envariance is invariance under simultaneous unitary transformations on the system and the environment. 
For a two state system entangled with an environment 
\begin{equation}
|\Psi \rangle =  c_1 | 1 \rangle | \epsilon_1 \rangle + c_2 | 2 \rangle | \epsilon_2 \rangle
\label{schmidt}
\end{equation}
and with $c_2 = c_1e^{i\phi}$, the transformation consists of two steps. 
First, a swap of the system states, $| 1 \rangle \leftrightarrow | 2 \rangle$, then a swap and phase change of the environment states. 
This combined unitary transformation, $U_e$,  will give
\begin{equation}
U_e|\Psi \rangle =  c_1 e^{i\phi}| 2 \rangle | \epsilon_2 \rangle + c_2 e^{-i\phi} | 1 \rangle | \epsilon_1 \rangle,
\label{swaped}
\end{equation}
so that the state is invariant under $U_e$. 

Zurek argues that envariance implies equal probability. They are all based on unproven assumptions and only the version in \cite{QDarwinPhysToday} will be discussed here. There, he argues by the following three steps. 
1) In state (\ref{schmidt}), the probability of the system being in state $| i \rangle$, $i=1,2$, is the same as the probability of the environment being in state $| \epsilon_i \rangle$. 
2) After the swap $| 1 \rangle \leftrightarrow | 2 \rangle$, the probability of the system being in state $1$ is has the same probability as $| \epsilon_2 \rangle$ and vice versa. 
3) The swap doesn't change the probability for $| \epsilon_1 \rangle$ or $| \epsilon_2 \rangle$.
From this follows that the probability of $| 1\rangle$ and $| 2 \rangle$ is the same, but the 1-3) are unproven assumptions.
They seem to be taken as `obvious' in spite of that it was not even explained how states get a probability associated to them.
 
Carroll and Sebens \cite{sebens2014self} argued on philosophical grounds for the ``Epistemic Separability Principle"  from which they proved Born's rule in two different ways \cite{sebens2014self,carroll2014many}. 
The principle is roughly stated as \cite{carroll2014many} `` the probability assigned post-measurement/pre-observation to an outcome of an experiment performed on a specific system shouldn’t depend on the physical state of other parts of the universe".
Their philosophical argumentation for their principle was based on the world we are experiencing, rather than from quantum mechanics. 
This means that they assumed that quantum mechanics give rise to the world we encounter. 
For this reason, their argumentation seems circular but we could instead consider the principle as an unproven assumption. 

Wallace has vigorously pursued to derive Born's rule in EQM. 
In his book \cite{wallace2012emergent} there are two different proofs. 
In the first, he assumed there is some function of subspaces of Hilbert space to a positive real number that fulfills what a probability reasonably should fulfill. 
Wallace implicitly assumed that unitary transformations don't change probabilities which imply that the probabilities are independent of context. He also assumed that if the state is an eigenstate, the eigenvalue has probability one. From this, he can convincingly derive that the probabilities equal $\rho_u$.  

Wallace second proof is based on decision theory. 
It does not presume the existence of probabilities but derives that in a fashion close Savage classical analysis \cite{savage1972foundations}. 
Wallace formulates ten axioms that his proof of Born's rule is based on. 
The proof is very complex and the steps of the reasoning are not sufficiently explained.
 
\section{Alternative EQM postulates \label{sec:altposts}}

The attempts to derive the Born rule from the relative frequency  \cite{hartle1968quantum,Farhi1989368,gutmann1995using} have started from a set of postulates corresponding to the Copenhagen postulates C1-3.
The postulate C4 that contains Born's rule have been replaced by C4'.

Wada \cite{wada2007derivation,wada2009minimal} has proposed to base EQM on a postulate equivalent to C4'  in which the density $\rho_u = 1$ replaces the requirement that $B| u \rangle = u | u \rangle$. 

The merits of these postulates are that C4' is very natural and easily acceptable and that the postulates are very much the same as the traditional Copenhagen postulates.  

The problems with defining EQM with C1-3 and C4' are substantial. 
So far, no proof of Born's rule exists. 
The statement in C1, that the state is a member of a Hilbert space, is a mathematical statement and it lacks direct physical content. 
In the absence of an interpretation of $\rho_{\cdot}(\cdot)$ as in EQM1 or Born' rule, the decoherence argument for the appearance of many worlds is difficult to make. 
Without EQM1 or something similar, any attempt to prove Born's rule is doomed to be circular as it would have to rely on decoherence to grant the appearance of randomness. Decoherence theory needs to founded on some interpretation of the quantum state that C1-C3 don't give. 

The very high-level assumptions used by Zurek \cite{zurek2003environment,BornRuleEnvariance,QDarwinPhysToday} and Sebens and Carroll  \cite{sebens2014self,carroll2014many}, which include the existence of probabilities, are not suitable to be reformulated into postulates. The probabilities that appear in EQM should rather be emergent, not postulated. 
Quantum probabilities are different from classical probabilities as in the quantum case all alternatives occur. They cannot be postulated in terms of classical probabilities, at least not within Everett's many worlds theory.

\section{Conclusions}

By formulating postulates for quantum mechanics that interpret the physical significance of the density $\rho_{\cdot}(\cdot) = |\psi_{\cdot}(\cdot)|^2$ Everett's vision of quantum mechanics has become a possible description of the physical world.

The proposed postulates make it possible to use the existing decoherence analysis without assuming Born's rule. This avoids the circularity of the arguments leading up to Born's rule. The statement that quantum states are Hilbert space vectors has been derived instead of postulated. By treating measurements as a physical process the measurement it has been derived that measurements correspond to a Hermitian operator and the measured values are the eigenvalues of the operator. 
It was also proven that in a typical world, physicists who have done many repeated experiments see events that are extremely reasonable to understand as probabilistic in accordance with Born's rule. This derivation of Born's rule should be acceptable from both bayesian and frequentist point of view. 

Previous attempts to prove Born's rule have been shown to be incomplete and the postulates used in this context are shown not be appropriate. The lack of physics content of the starting points is the main reason for troubles encountered.

Everett's vision is completed even though it was may be necessary to deviate from the fundamental point of interpretation. Everett \cite{everett1957relative} wrote, "... the theory itself sets the framework for its interpretation".  This statement is correct, but Everett and many others seem to have concluded that no interpretation of the equation was needed. What applies it that the interpretation has to be in accordance with both the logical structure of the theory and with the physical world. This has been supplied here by postulate EQM1.

The present study has shown that the measurement process is fully described by quantum mechanics. 
The invisibility of the measurement problem in the practical analysis of quantum systems is explained by that there are no other
mechanisms than that of quantum unitary evolution. It turns out that the idea that there is no mechanism responsible for the transition from uncertainty to certainty \cite{bohr2004principle} is partly true. There is no extra mechanism only quantum mechanics.
Previously, the appearance of the macroscopic has been understood in quantum mechanics due to the effects of decoherence  \cite{joos2013decoherence,zurek2003decoherence,schlosshauer2007decoherence}.
 
It is now possible to conclude that there is no known type of situation in which a transition into a classical particle occur in a fundamental way rather than in an emergent or semi-classical way.

\section{Final remarks}

In pedagogical connections, the term probability amplitude is not appropriate except in historical accounts. 
The density $\rho_{\cdot}(\cdot)$ should simply be called density rather than probability density except in connection to actual measurement processes. The traditional language suggests that there is a real classical point particle that is to be found with some probability, but there are no scientific reasons for such a view.

Though all particles are waves, it is still appropriate to use the term particle. The Ehrenfest theorem reinforced by decoherence shows that a wave packet with classical particle properties appears, given the right circumstances. Scattering between free particles and deep inelastic scattering on atoms, nuclei, and nucleons show similarities with classical particle collisions. 

Objections have been raised \cite{SquiresOnProof,cassinello1996probabilistic} against previous work \cite{hartle1968quantum,Farhi1989368} that, by the use of the Hilbert space norm or scalar product, the Born rule is already assumed. The fact that the norm can be viewed to be a measure doesn't imply by itself that it is a probability measure. This is stated to counter similar objections towards the postulate EQM1. All distributions are not necessarily probability distributions. The $\rho_{\cdot}(\cdot) = |\psi_{\cdot}(\cdot)|^2 $ is a distribution that gives the distributed position of the system. That this implies $\rho$ also to define probabilities is not a trivial result as the history proves.

There is no claim that the quantum state is real (ontological), only that quantum physics describe nature as far as we have encountered it including the measurement process. The author hopes that readers will analyze the presented theory as a physics theory, without any prejudice about how the world looks like. Is the proposed theory a description of nature as we know it?

\begin{acknowledgments}
I wish to acknowledge Ben Mottelson for useful and interesting discussions on the foundation of quantum mechanics.
\end{acknowledgments}

\appendix*

\section{HILBERT SPACES}

I wish to acknowledge Ben Mottelson for trying under several decades to make me work on quantum foundations. Finally, I am on it.
A Hilbert space $H$ is a normed linear space \cite{reed1980methods}. The members are often called vectors. The norm $\|\cdot \|$ attributes a real number $\geq 0$ for any member in $H$. The normed space has the property that
\begin{equation}
\| x \| = 0 \Leftrightarrow x=0.
\end{equation}
This implies that
\begin{equation}
\| x - x' \| = 0 \Leftrightarrow x=x'.
\label{equivalence}
\end{equation}
An example, the equation for an eigenvalue $a$ of a linear operator $A$ is
\begin{equation}
Ax=ax \Leftrightarrow  \| Ax - ax \| = 0.
\end{equation}
For a Hilbert space consisting of functions ($\mathbb{R}^n \rightarrow \mathbb{C}, n\in \mathbb{N}$ or $n=\infty$) the functions make up equivalence classes. 
Two functions $\psi$ and $\psi'$ belong to the same equivalence class if they are equal almost everywhere in the sense $\| \psi - \psi' \| = 0$. 
The vectors of the Hilbert space are the equivalence classes. 
If it is clear that we deal with vectors in Hilbert space, then the statement $\psi  = \psi'$ means $\| \psi - \psi' \| = 0$. 

The most important property of Hilbert spaces is the existence of an inner product $\langle x | y \rangle $ which is related to the norm by $\|x\|^2 = \langle x |x \rangle$. Furthermore, $\langle x | y + \lambda z \rangle = \langle x | y \rangle + \lambda \langle x | z \rangle$ and $\langle y | x \rangle = \langle x | y \rangle^*$.

Hilbert spaces are complete and spaces with an inner product can be completed to become a Hilbert space. Any Hilbert space has a complete orthonormal basis set. 

In separable Hilbert spaces, the basis set is countable. 
Then any vector $\psi$ can be written as a sum of the basis states $\phi_b$,
\begin{equation}
\psi = \sum_b \langle \phi_b | \psi \rangle \phi_b
\label{expansion}
\end{equation}
This equation holds only in the sense of the equivalence (\ref{equivalence}).  If $\|\psi-\psi'\| = 0,\,$ then the matrix elements $\langle \phi_b | \psi \rangle = \langle \phi_b | \psi' \rangle$. Both $\psi$ and $\psi'$  give rise to the same lefthand side.

The Hilbert space for a finite number of particles is separable.
The Hilbert space for infinitely many particles $H^{\otimes \infty}= H\otimes H \otimes ...$, which is an infinite tensor product of separable Hilbert spaces. 
This space is not separable, so any complete basis set will be uncountable. This can result in that the inner product $\langle \phi | \psi \rangle$ between normalized basis states $\phi$ and the normalized state $\psi$ all are zero. At most a countable set of inner products can be non-zero.

The left hand side of (\ref{norm}) defines a norm and there is a unique inner product 
\begin{eqnarray}
 \lefteqn{\langle \Psi | \Phi \rangle =  } \nonumber\\
 & \int\! \int \!\cdots dx_1dx_2 \cdots \sum_j \psi_j(t, \mathbf x_1,\mathbf x_2, \ldots)^*
 	 \phi_j(t, \mathbf x_1,\mathbf x_2, \ldots).  \nonumber\\
\end{eqnarray}
 
As stated in the comment to EQM1, there is no observable difference between $\Psi$ and $\Psi'$ if $\|\Psi - \Psi'\| =0 $. 
This shows that all (observable) physical properties of the system are represented by the vector in Hilbert space. 

\bibliography{../Bib_physics}

\end{document}